\documentclass[aps,prb,showpacs,tightenlines,twocolumn,nofootinbib,nobibnotes,superscriptaddress]{revtex4-1}
\usepackage{amsmath,amssymb,amsfonts,bm}
\usepackage{graphicx}
\usepackage{epstopdf}
\usepackage{dcolumn}
\usepackage{mathrsfs}
\usepackage{bbold}
\usepackage{dsfont}
\usepackage{dcolumn}

\usepackage[colorlinks=true,linkcolor=blue,citecolor=blue, urlcolor=blue,bookmarks=false]{hyperref}

\begin{document}

\title{Topological Anderson insulator phase in a quasicrystal lattice}
\date{\today }

\author{Rui Chen}

\author{Dong-Hui Xu}\thanks{donghuixu@hubu.edu.cn}

\author{Bin Zhou}\thanks{binzhou@hubu.edu.cn}
\affiliation{Department of Physics, Hubei University, Wuhan 430062, China}
\begin{abstract}
Motivated by the recent experimental realization of the topological Anderson insulator and research interest on the topological quasicrystal lattices, we investigate the effects of disorder on topological properties of a two-dimensional Penrose-type quasicrystal lattice that supports the quantum spin Hall insulator~(QSHI) and normal insulator~(NI) phases in the clean limit. It is shown that the helical edge state of the QSHI phase is robust against weak disorder. Most saliently, it is found that disorder can induce a phase transition from NI to QSHI phase in the quasicrystal system. The numerical results based on a two-terminal device show that a quantized conductance plateau can arise inside the energy gap of NI phase for moderate Anderson disorder strength. Further, it is confirmed that the local current distributions of the disorder-induced quantized conductance plateau are located on the two edges of sample. Finally, we identify this disorder-induced phase as topological Anderson insulator phase by computing the disorder-averaged spin Bott index.
\end{abstract}

\maketitle

\section{Introduction}
The quasicrystal, which possesses long-range orientational order but non translational symmetry~\cite{Shechtman1984PRL,Levine1984PRL,Biggs1990PRL,Pierce1993Science,Basov1994PRL,Pierce1994PRL}, exhibits various unusual physical properties such as extremely low friction~\cite{Dubois1991JMSL}, self-similarity~\cite{Sutherland1986PRB,Fujiwara1988PRB}, and critical behaviours of wave functions~\cite{Tsunetsugu1991PRB1,Tsunetsugu1991PRB2}. Moreover, the quasicrystal possesses multifractal energy structure~\cite{Fujiwara1989PRB}, which is responsible for the counterintuitive transport properties, including the abnormally low conductivities~\cite{Fujiwara1989PRB,Matsubara1991JJAP} and the disorder-enhanced transport~\cite{Fujiwara1993PRL,Mayou1993PRL} that was experimentally confirmed in a photonic system~\cite{Levi2011Science}. Due to its unique properties, the quasicrystal has attracted much attention theoretically and experimentally, and has been investigated in various systems including the solid state systems~\cite{jenks1998Langmuir,McGrath2002JPCM,Sharma2007AdvPhy,McGrath2012PTRSA}, photonic systems~\cite{Vardeny2013NatPho,Viebahn2019PRL,Steurer2007JPD}, and phononic systems~\cite{Steurer2007JPD,Chen2008SSC}.

Materials with nontrivial topology have attracted much attention in the last decade~\cite{Qi2011RMP,Hasan2010RMP,Bansil2016RMP}. Nowadays, the search for new topological phases becomes a fundamental theme of the condensed matter physics community. Topological materials were mainly investigated in conventional crystalline systems. Recently, an increasing interest is realizing nontrivial topological phases in the quasicrystal systems. For example, Huang and Liu proposed that the quantum spin Hall insulator (QSHI) can be realized on a Penrose-type quasicrystal lattice, in which the bulk topology is captured by the spin Bott index~\cite{Huang2018PRL,Huang2018PRB}. Furthermore, they found that the QSHIs manifest similarly in an Ammann-Beenker-type octagonal quasicrystal and a periodic snub-square crystal, indicating the robustness of the topological properties regardless of symmetry and periodicity~\cite{Huang2019PRB}. Moreover, the quantum Hall insulator~\cite{Tran2015PRB}, the Floquet Chern insulator~\cite{Bandres2016PRX}, the topological superconductor~\cite{Poyhonen2018NatCom,Fulga2016PRL}, the non-Hermitian topological insulator~\cite{longhi2019arXiv,zeng2019arXiv}, and the higher-order topological insulator~\cite{Chen2019arXivHigh}/superconductor~\cite{DanielVarjas2019arXiv} were proposed in the quasicrystal systems. In the mean time, the topological phase transitions were experimentally observed in photonic quasicrystals~\cite{Verbin2013PRL}, and recently the topological edge modes were also reported in quasiperiodic acoustic waveguides~\cite{Apigo2019PRL}.

On the other hand, the interplay between topology and disorder plays an important role in the recent research of topological matters. The topological Anderson insulator (TAI), one of disorder-induced topologically nontrivial phases, was firstly proposed by Li \emph{et al} ~\cite{Li2009PRL}. Since then, the TAI and the disorder effects on the topological matters have been investigated in various models and systems ~\cite{Jiang2009PRB,Groth2009PRL,Wu2016CPB,Xing2011PRB,Orth2016Scirep,GUO2010PRL,Guo2011PRB,Guo2010PRB,Chen2017PRBDirac,Chen2017PRBLieb,Chen2018PRBWeyl,Chen2018PRBWeylFloquet,
Su2016PRB,Kimme2016PRB,ChenCz2015PRL,ChenCZ2015PRB,Shapourian2016PRB,Kuno2019PRB,Zhang2019arXiv}. Recently, the TAI phase has been observed experimentally in one-dimensional disordered atomic wires~\cite{Meier2018Science} and a photonic platform~\cite{Stutzer2018Nature}. It is noted that up to now the studies of the TAI phase mainly focus on the crystalline systems with disorder. Considering the recent processes of topological nature of quasicrystal systems, thus, the disorder effects on topological quasicrystal systems is also an interesting issue worthy to be addressed.

In this work, we investigate the disorder effects on a general atomic-basis quasicrystal lattice model arranged according to the Penrose tiling. By calculating the two-terminal conductance, it is found that disorder can induces a quantized conductance plateau with $G=2e^2/h$ from the normal insulator (NI) phase. The quantized plateau has zero conductance fluctuation, and maintains in a certain range of disorder strength. Thus, it is indicted that the TAI phase appears in the quasicrystal system. Furthermore, the inspection of the nonequilibrium local current distribution confirms that the quantized conductance plateau arises from the disorder-induced helical edge states. Finally, the TAI phase can be characterized by the quantized disorder-averaged spin Bott index.

\section{Model and method}

\begin{figure}[ptb]
\includegraphics[width=8cm]{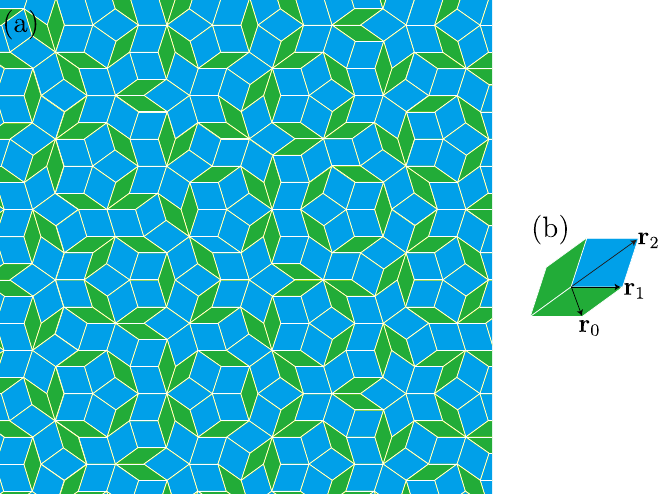}
\caption{(a) Schematic illustration of the Penrose tiling, which is composed of the fat and thin rhombuses that can tile the 2D plane in an aperiodic way. (b) The first three NN hoppings in the Penrose-type quasicrystal lattice with the distance ratio being $r_0:r_1:r_2=2\cos\frac{2\pi}{5}:1:
2\sin\frac{\pi}{5}.$}
\label{fig1}
\end{figure}

Following the pioneer works by Huang and Liu~\cite{Huang2018PRL,Huang2018PRB}, we start with a two-dimensional (2D) quasicrystal lattice with its sites arranged on each vertex of the Penrose tiling~\cite{Penrose1974BIMA}, as depicted in Fig.~\ref{fig1}(a). The Penrose tiling contains two types (thin and fat) of rhombuses, which can fill the plane completely in an aperiodic way. To construct the Penrose tiling, we adopt Bruijn's method by projecting a five-dimensional cubic lattice onto the 2D plane~\cite{Bruijn1981IM}. The QSHI model on the Penrose-type quasicrystal lattice introduced by Huang and Liu~\cite{Huang2018PRL,Huang2018PRB} is described by the following Hamiltonian,
 \begin{eqnarray}
H &=&\sum_{i\alpha }\epsilon _{\alpha }\sigma _{0}c_{i\alpha }^{\dag
}c_{i\alpha }+\sum_{\left\langle i\alpha ,j\beta \right\rangle }t_{i\alpha
,j\beta }\sigma _{0}c_{i\alpha }^{\dag }c_{j\beta }  \nonumber \\
&&+i\lambda \sum_{i}\left( c_{ip_{y}}^{\dag }\sigma
_{z}c_{ip_{x}}-c_{ip_{x}}^{\dag }\sigma _{z}c_{ip_{y}}\right) ,
\label{Hamiltonian1}
\end{eqnarray}%
where $c_{i\alpha }^{\dag }$ and $c_{i\alpha }$ are the creation and
annihilation operators of electrons with the orbital $\alpha =s,p_{x},p_{y}$
on site $i$. $\sigma _{i}$ are Pauli matrices representing spin. $\epsilon
_{\alpha }$ is the magnitude of the onsite energy of the $\alpha$-orbit.
The second term corresponds to the hopping integral with its amplitude
depending on the orbital type and the distance vector $\mathbf{d}_{ij}=\mathbf{%
r}_{j}-\mathbf{r}_{i}.$ The last term is the spin-orbit coupling (SOC) term,
and $\lambda $ corresponds to the SOC strength. In this model, the hopping integral $%
t_{i\alpha ,j\beta }$ is determined by the Slater-Koster parametrization~\cite{Slater1954PR}
\begin{equation}
t_{i\alpha ,j\beta }\left( \mathbf{d}_{ij}\right) =\text{SK}\left[ V_{\alpha \beta \gamma}\left( d_{ij}\right) ,\mathbf{\hat{d}}_{ij}\right] ,
\end{equation}%
which depends on the orbital type $\alpha,\beta=s,p_x,p_y$, the vector $\mathbf{d}_{ij}$, and the bonding parameter $\gamma=\sigma,\pi$. Here the specific form of $t_{i\alpha ,j\beta }\left( \mathbf{d}_{ij}\right) $ is presented in Table \ref{tab1}, where $\mathbf{\hat{d}}_{ij}=\left( l,m\right) $ is the unit direction
vector along $\mathbf{d}_{ij}$, and $d_{ij}=\left|\mathbf{d}_{ij}\right|$ is the distance between $i$-th and $j$-th sites. The bonding parameters $V_{\alpha \beta \gamma}\left( d_{ij}\right)$ depend on the distance between two sites, and can be expressed by the Harrison relation~\cite{Harrison2012book}
\begin{equation}
V_{\alpha \beta \gamma}\left(d_{ij}\right)
=V_{\alpha \beta \gamma}^0\frac{d^2_0}{d^2_{ij}},
\end{equation}
where $d_0$ modulates the overall bonding amplitudes. In the subsequent calculations, we take the parameters as $\epsilon_s=1.8 \text{ eV}$, $\epsilon_{p_x}=\epsilon_{p_y}=-6.5 \text{ eV}$, $\lambda=1 \text{ eV}$, $V_{ss\sigma}^0=-0.4 \text{ eV}$, $V_{sp\sigma}^0=0.9 \text{ eV}$, $V_{pp\sigma}^0=1.8 \text{ eV}$, and $V_{pp\pi}^0=0.05 \text{ eV}$. We will only consider the nearest-neighbor (NN) hopping, the next-nearest-neighbor hopping, and the next-next-nearest-neighbor hopping, which correspond to the short diagonal of the thin rhombuses, the edge of rhombuses, and the short diagonal of the fat rhombuses, respectively, shown in Fig.~\ref{fig1}(b). In the previous works by Huang and Liu~\cite{Huang2018PRL,Huang2018PRB,Huang2019PRB}, based on the Penrose-type
quasicrystal lattice model described by the Hamiltonian (\ref{Hamiltonian1}), they proposed that the quantum spin Hall effect can be realized in quasicrystal lattices.

\begin{table}[t]
\begin{ruledtabular}
\caption{The Slater-Koster hopping matrix elements.}
\begin{tabular}{c|ccc}
Orb.&   $s$&   $p_x$&$p_y$
\\\hline
$s$&$V_{ss\sigma}$&$lV_{sp\sigma}$&$mV_{sp\sigma}$
\\
$p_x$     &$-lV_{sp\sigma}$&$l^2V_{pp\sigma}+\left(1-l^2\right)
V_{pp\pi}$
&$lm\left(V_{pp\sigma}-V_{pp\pi}\right)$
\\
$p_y$&$-mV_{sp\sigma}$&
$lm\left(V_{pp\sigma}-V_{pp\pi}\right)$     &$m^2V_{pp\sigma}+\left(1-m^2\right)V_{pp\pi}$
\\
\end{tabular}
\label{tab1}
\end{ruledtabular}
\end{table}

We will investigate the transport properties of the Penrose-type quasicrystal lattice with disorder by using the
Landauer-B\"uttiker-Fisher-Lee formula
\cite{Landauer1970Philosophical,Buttiker1988PRB,
Fisher1981PRB} and the recursive Green's
function method \cite{Mackinnon1985Zeitschrift,Metalidis2005PRB}. The linear conductance can be obtained by $G=(e^2/h)T$, where $T=$ Tr$\left[ \Gamma_L G^r \Gamma_R G^a \right]$ is the transmission coefficient. The linewidth function $\Gamma_{\eta}(\mu)=i\left[ \Sigma_{\eta}^{r}-\Sigma_{\eta}^{a}\right]$ with $\eta=L,R$, and the Green's functions $G^{r/a}(\mu)$ are calculated from $G^{r}(\mu)=\left[ G^a(\mu)\right]^\dag=\left[\mu I-H_C-\Sigma_{L}^{r}-\Sigma_{R}^{r} \right]^{-1}$, where $\mu$ is the chemical potential, $H_C$ is the Hamiltonian matrix of the central scattering region, and $\Sigma_{L,R}^{(r/a)}$ are the retarded (advanced) self-energy due to the device leads.

\begin{figure}[ptb]
\includegraphics[width=8cm]{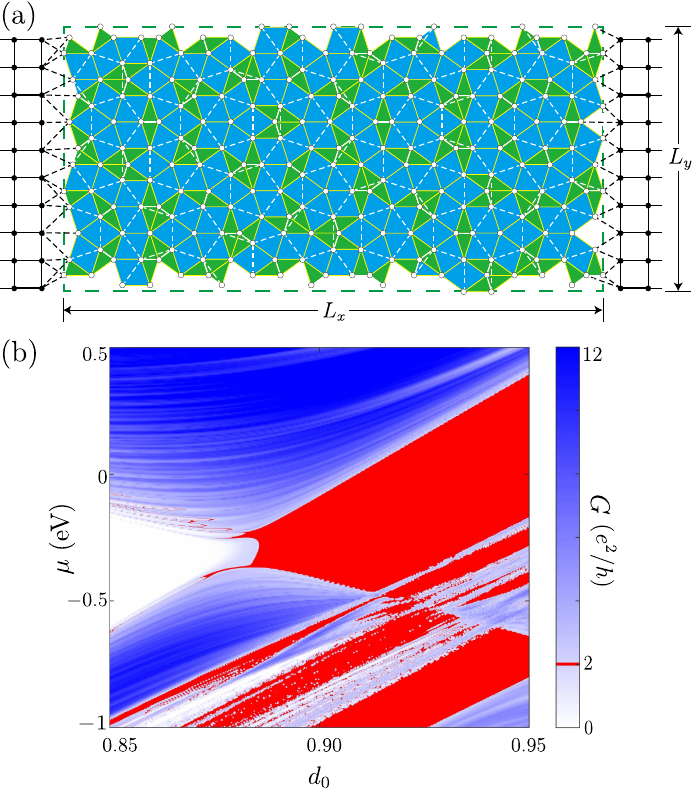}
\caption{(a) Schematic illustration of the two-terminal device used in the transport simulations for the Penrose lattice. (b) Phase diagram for the conductance as functions of the chemical potential $\mu$ and $d_0$ in the clean limit. In our numerical simulations, the width and length of the sample are chosen to be $L_x=200 r_1$ and $L_y=100 r_1$.}
\label{fig2}
\end{figure}

As shown in Fig.~\ref{fig2}(a), we adopt a clipped Penrose lattice of size $L_x\times L_y$ as the central scattering device, and two semi-infinite square lattice as the leads connected to the device along the $x$ direction. The Hamiltonian for the semi-infinite lead is
\begin{equation}
H_L= \sum_{i\alpha }\mu_L\sigma _{0}c_{i\alpha }^{\dag
}c_{i\alpha }+ \sum_{\left\langle i\alpha ,j\beta \right\rangle }t_{i\alpha
,j\beta }\sigma _{0}c_{i\alpha }^{\dag }c_{j\beta },
\end{equation}
where $\mu_L$ is the chemical potential of the leads. In modeling the leads, we only consider the NN hopping of the square lattice, i.e., $d_{ij}=r_1$. The Hamiltonian that describes the connection between the leads and device is
\begin{equation}
H_{LD}=\sum_{\left\langle i_s\alpha ,j_s\beta \right\rangle }t_{i_s\alpha
,j_s\beta }\sigma _{0}c_{i_s\alpha }^{\dag }c_{j_s\beta },
\end{equation}
where the labels $i_s$ and $j_s$ correspond to the connected points in the leads and device, respectively. In the subsequent calculation, we take $\mu_L=2 \text{ eV}$ to guarantee a high density of state of the leads. We will introduce the Anderson-type disorder to the central scattering region with
\begin{equation}
\Delta H=\sum_{i\alpha}{W_{i\alpha}\sigma_{0}c_{i\alpha }^{\dag
}c_{i\alpha }},
\end{equation}
where $W_{i\alpha}$ is uniformly distributed within $\left[-U/2,U/2\right]$, with $U$ being the disorder strength.

On the other hand, the Penrose quasicrystal lacks the translational symmetry, and cannot be analysed in the framework of Bloch theory. However, the quasicrystal tiling approximants provides a systematic approach to construct a periodic lattice which can gradually approximate the aperiodic quasicrystal with increasing the vertex number of the super unit cell~\cite{Tsunetsugu1991PRB1,Tsunetsugu1991PRB2,
Tsunetsugu1986JPSP,Entin_Wohlman1988JP}. It is noted that the spin Bott index has been proposed to identify the quantum spin Hall state in both crystalline and nonperiodic systems~\cite{Huang2018PRL,Huang2018PRB}. Thus, in the Penrose-type quasicrystal lattice, we will adopt the spin Bott index to identify the topological phase. To calculate the spin Bott index in real space, one needs to construct the projector operator of the occupied states as
\begin{equation}
P=\sum_{i}^{N}\left\vert \psi_{i}\right\rangle \left\langle \psi
_{i}\right\vert ,
\end{equation}
where $\psi_{i}$ is the $i$-th eigenvector of the Hamiltonian obtained by
the Penrose-type quasicrystal tiling approximants, and $N$ is the number of the occupied
bands. Then we construct the projected spin operator as $P_{z}=P\hat{s}_{z}P$,
where $\hat{s}_{z}=\frac{\hbar}{2}\sigma_{z}$ is the spin operator. The
eigenvalue of $P_{z}$ is composed of two parts that are separated by zero-value energy. The
number of positive eigenvalues is $N/2$, which equals to the number of
negative eigenvalues. In this way, the new projector operators can be
constructed as
$P_{\pm}=\sum_{i}^{N/2}\left\vert \phi_{i}^{\pm}\right\rangle \left\langle
\phi_{i}^{\pm}\right\vert$,
where $\phi_{i}^{+}(\phi_{i}^{-})$ is the eigenvector of the $i$-th positive
(negative) eigenvalue. One defines the projected position operators of the two
spin sectors as%
\begin{align}
U_{\pm}  & =P_{\pm}e^{i2\pi X/L_{x}}P_{\pm}+\left(  I-P_{\pm}\right) \nonumber ,\\
V_{\pm}  & =P_{\pm}e^{i2\pi Y/L_{y}}P_{\pm}+\left(  I-P_{\pm}\right)
,
\end{align}
where $X_{ii}=x_{i}$ and $Y_{ii}=y_{i}$ are diagonal matrices, and $\left(
x_{i},y_{i}\right)  $ is the coordinate of the $i$-th lattice. In order to make the
results more stable, we adopt the singular value decomposition $M=Z\Lambda
\Pi^{\dag}$ for $U_{\pm}$ and $V_{\pm},$ and then set $\tilde{M}=Z\Pi^{\dag}$
as the new unitary operators. Finally, the spin Bott index is obtained as~\cite{Huang2018PRL,Huang2018PRB}%
\begin{equation}
B_{s}=\frac{1}{2}\left(  B_{+}-B_{-}\right)  ,
\end{equation}
with $B_{\pm}=\frac{1}{2\pi}\operatorname{Im}\left[  \operatorname{Tr}\left(  \ln V_{\pm
}U_{\pm}V_{\pm}^{\dag}U_{\pm}^{\dag}\right)  \right]  $ being the Bott indexes for the two spin sectors.

\section{Numerical simulation}
\subsection{The clean limit}
In Fig.~\ref{fig2}(b), we compute the conductance as a function of the chemical potential $\mu$ and $d_0$ for the system being in the clean limit, i.e., $U=0$. Firstly, let us focus on the $\frac{2}{3}$ filling of the electron states, which corresponds to the chemical potential is about $\mu=-0.3\text{ eV}$. From Fig.~\ref{fig2}(b), it is observed that the conductance is zero for $d_0$ being less than $0.88$ and $\mu=-0.3\text{ eV}$. Thus, in this case the system is a NI. While for a larger $d_0$, the conductance shows a quantized value of $G=2e^2/h$, implying the nontrivial topological nature of the system. On the other hand, when the chemical potential is located in the bulk bands, the multifractal band structure~\cite{Fujiwara1989PRB} can be clearly observed from the phase diagram [Fig.~\ref{fig2}(b)], that is each band of the quasicrystal system has several pseudogaps that separate it to subbands, which leads to the regions with zero or quantized conductances in the phase diagram. In this work, however, we will only concentrate on the gap near $\frac{2}{3}$ filling of electron states.

By calculating the spin Bott index on the approximated Penrose quasicrystal system, we confirm that for $d_0>0.88$ the system is a QSHI which is characterized by nonzero spin Bott index $B_s=1$ (corresponding to the quantized conductance $G=2e^2/h$). While for $d_0<0.88$, the system is a NI with $B_s=0$.

\subsection{The disorder effects}

\begin{figure}[tpb]
\includegraphics[width=8.5cm]{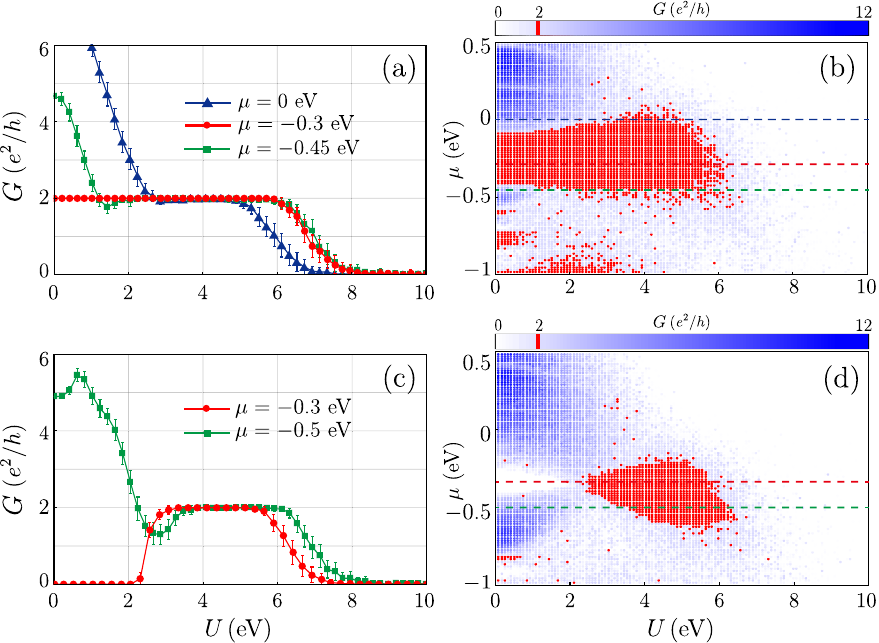}
\caption{(a) The conductance as a function of the disorder strength $U$ for $d_0=0.90$ with different chemical potentials $\mu=0\text{ eV}$, $\mu=-0.3 \text{ eV}$, and $\mu=-0.45 \text{ eV}$, which correspond to the blue, red, and green dashed lines in (b), respectively. The error bars show the standard deviation of the conductance for 1000 samples. (b) Phase diagram for the conductance as functions of the disorder strength $U$ and the chemical potential $\mu$. In our numerical simulations, the width and length of the sample are chosen to be $L_x=200 r_1$ and $L_y=100 r_1$, respectively. (c-d) The same as (a-b), except that $d_0=0.87$.
}
\label{fig3}
\end{figure}

Now, we study the disorder effects on the 2D quasicrystal lattice. In Figs.~\ref{fig3}(a) and \ref{fig3}(c), we respectively plot the conductance as a function of the disorder strength $U$ for $d_0=0.90$ and $d_0=0.87$ with several different values of chemical potential $\mu$. For the case of $d_0=0.90$ [Fig. ~\ref{fig3}(a)], in the clean limit, the system is a QSHI with a quantized conductance $G=2e^2/h$ when the chemical potential lies in the bulk energy gap ($\mu=-0.3 \text{ eV}$). With increasing disorder strength, the conductance keeps the quantized value until the disorder strength $U$ exceeds $6\text{ eV}$. Therefore, similar to the previous studies on disordered crystalline systems~\cite{Li2009PRL,Jiang2009PRB,Groth2009PRL,Wu2016CPB}, the topologically protected helical edge states in the Penrose-type quasicrystal lattice are also robust against disorder. When the chemical potential is located at the bulk bands ($\mu=0 \text{ eV}$ and $\mu=-0.45 \text{ eV}$), the conductance is suppressed by disorder, then gradually decreases to the quantized plateau before it finally disappears.

While for a NI phase with $d_0=0.87$ and $\mu=-0.3 \text{ eV}$ [as shown by red curve in Fig. ~\ref{fig3}(c)], the conductance is zero as no state appears inside the bulk energy gap when the disorder strength is weak. With increasing disorder strength ($U>2\text{ eV}$), the conductance increases and then forms a quantized plateau with $G=2e^2/h$. The quantized plateau is observed for a certain range of disorder strength, and it decreases and finally disappears with increasing the disorder strength. The zero conductance fluctuation of the quantized plateau indicates that it may be originated from the helical edge state of the QSHI phase. The case with $d_0=0.87$ and $\mu=-0.5 \text{ eV}$ is also presented in Fig.~\ref{fig3}(c) [as shown by green curve]. It is observed that in this case the variation of conductance with increasing disorder strength is similar to that of the case with $d_0=0.90$ and $\mu=0 \text{ eV}$ (or $\mu=-0.45 \text{ eV}$) shown in Fig. ~\ref{fig3}(a).

In Figs.~\ref{fig3}(b) and \ref{fig3}(d), we plot the phase diagrams of the QSHI state and the NI state on the $(U,\mu)$ phase space, respectively. Here each point is obtained from a single configuration of disorder, which is enough to determine the quantized conductance region. It can be clearly shown that the disorder-induced quantized conductance region for the NI phase appears mainly in the energy gap, and the small region of the valence bands [Fig.~\ref{fig3}(d)].

To further confirm the assertion that the quantized conductance plateaus originate from the robust edge states, we calculate the nonequilibrium local
current distribution between sites $\mathbf{{i}}$ and $\mathbf{{j}}$ from the following formula\cite{Jiang2009PRB}
\begin{equation}
J_{\mathbf{{i}\rightarrow{j}}}=\frac{2e^{2}}{h}\text{Im} \left[  \sum
_{\alpha,\beta}{H_{\mathbf{{i} \alpha,{j}\beta}}G^{n}_{\mathbf{{j}\beta
			,{i}\alpha}}}\right]  \left(  V_{L}-V_{R} \right)  \text{,}%
\end{equation}
where $V_{L}(V_R)$ describes the voltage of the left (right) lead, and
$G^{n}=G^{r}\Gamma_{L} G^{a}$ is the electron correlation function. To calculate the local current distribution, a small external bias $V=V_L-V_R$ is applied longitudinally between the two terminals, where $V_L$ and $V_R$ describe the voltages of the left and right leads. The small bias voltage $V$ is fixed to be $0.0001 \text{ eV}$. We assume the electrostatic potential in the central part is $\phi(x_i)=(L_x-x_i+1)V/(L_x+1)$, where $x_i$ is $x$ direction coordinate of the $i$-th site and $1\leq x_i \leq L_x$. Then, the electric field is uniformly distributed in the central sample region.

\begin{figure}[t]
\includegraphics[width=8.5cm]{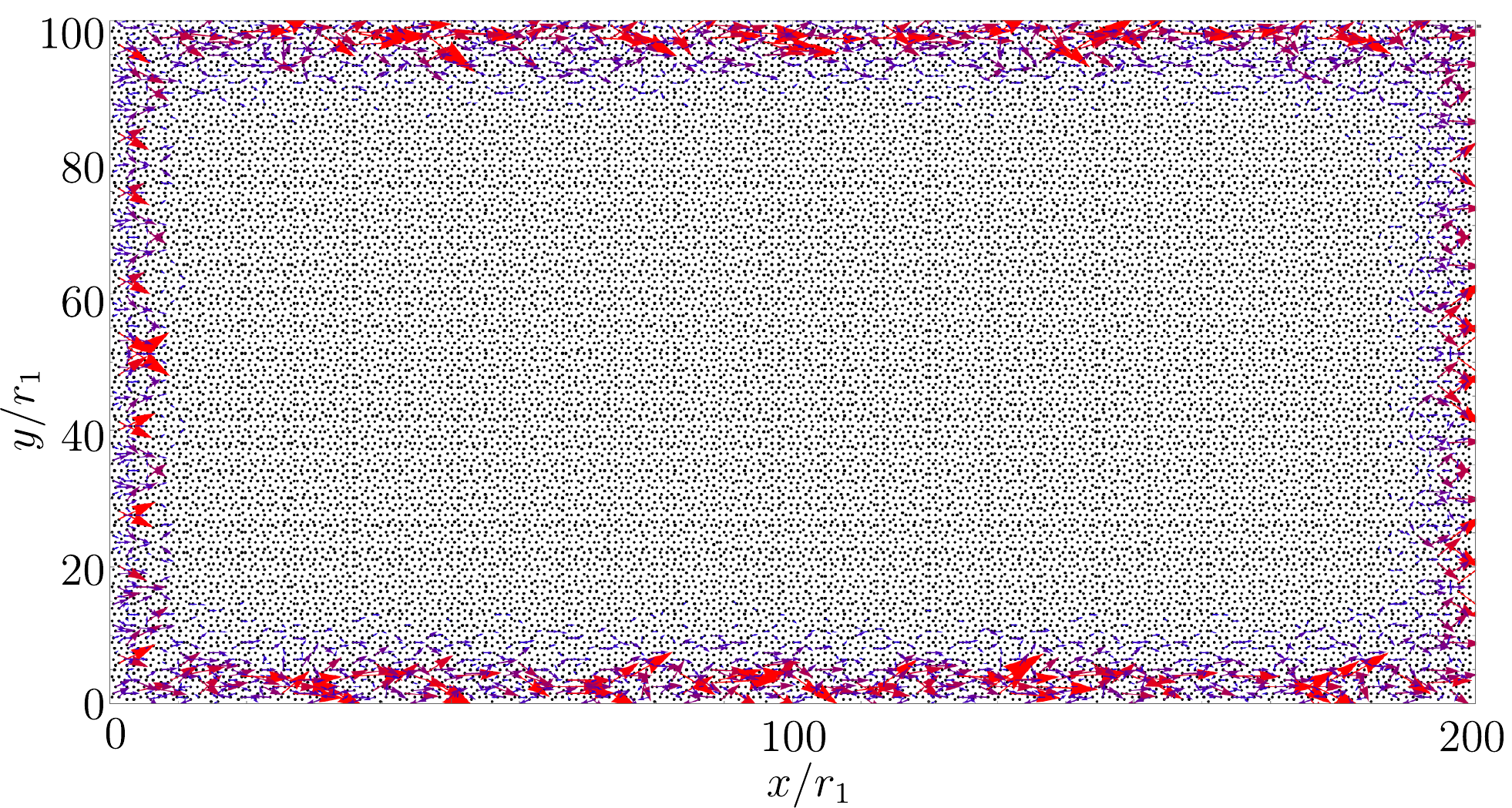}
\caption{The averaged nonequilibrium local current distribution of the disorder-induced QSHI phase on the Penrose lattice with $d_0=0.87$, $\mu=-0.3\text{ eV}$, and $U=4 \text{ eV}$. The arrow size means the current strength.
}
\label{fig4}
\end{figure}

In Fig.~\ref{fig4}, we plot the averaged current distribution~\cite{Jiang2009PRB} of the disorder-induced QSHI state. The size of the arrows corresponds to the local current strength. Apparently, for the disorder-induced QSHI phase with $G=2e^2/h$, the currents are localized at the two opposite sides of the sample, which characterizes the helical edge states of the disorder-induced QSHI phase.

\begin{figure}[ptp]
\includegraphics[width=8.5cm]{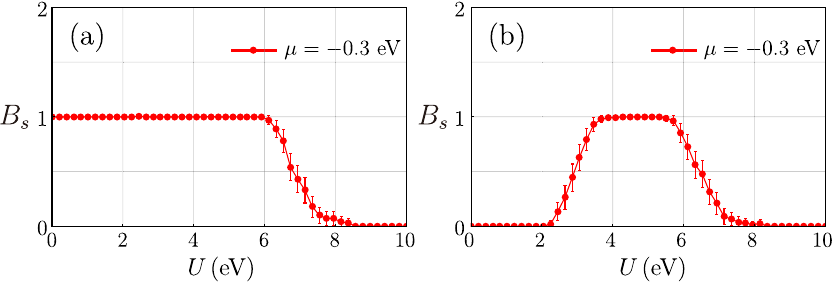}
\caption{Disorder-averaged spin Bott index on the approximated Penrose lattice with (a) $d_0=0.87$ and (b) $d_0=0.90$. The error bars show the standard deviation of the spin Bott index for 1000 samples.}
\label{fig5}
\end{figure}

In addition, the spin Bott index can also characterize the topological properties of the disordered system~\cite{Huang2018PRL,Huang2018PRB}. As shown in Fig.~\ref{fig5}, we plot the spin Bott index varying as functions of the disorder strength $U$ with $d_0=0.87$ and $d_0=0.90$. For the QSHI phase [Fig.~\ref{fig5}(a)], the spin Bott index is quantized as $1$, and maintains this value for a certain range of disorder strength. For the NI phase [Fig.~\ref{fig5}(b)], the spin Bott index is zero when the disorder is weak, and increases to the quantized value $1$ with increasing disorder strength. Therefore, we identify this disorder-induced topologically nontrivial phase as a TAI phase.

\section{Conclusion}
In this work, we study the effects of disorder on topological properties of a 2D Penrose-type quasicrystal lattice. Firstly, it is shown that the QSHI phase in the quasicrystal system is robust against weak disorder. Moreover, the disorder-induced TAI phase on a 2D quasicrystal lattice is proposed. The TAI phase is characterized by a disorder-induced quantized conductance plateau with $G=2e^2/h$. We also present the nonequilibrium local current distribution, and the disorder-averaged spin Bott index, which further identify the TAI phase. In the previous studies, the TAI phases were mainly presented in the crystalline systems~\cite{Li2009PRL,Jiang2009PRB,Groth2009PRL,Wu2016CPB}, while we extend the territory of TAI phase to the quasicrystal systems.

In conclusion, we found that the occurrence of TAI phase shares various similarities between the crystalline and quasicrystalline systems, for example, they both require that the NI phase is in proximity to the QSHI phase in the clean limit and the chemical potential is located near the band edge. The present results show that the disorder-induced phase transition is irrelevant to the lattice structure, and only depends on the above-mentioned conditions. In previous works on crystalline systems~\cite{Li2009PRL,Jiang2009PRB,Groth2009PRL,Wu2016CPB}, the disorder-induced topological phase transitions can be explained by a $k$-space self-consistent Born approximation, where disorder renormalizes the mass term and chemical potential, resulting in the TAI states. However, quasicrystal lacks the translational symmetry, and the original theory for the TAI phase is not available. Considering the similarities mentioned above, we could conjecture that the underlying reason for the occurrence of TAI phase in the Penrose-type quasicrystal lattice may be attributed to the renormalization effect of disorder, and it will be investigated in our future works.

Recently, the TAI phases in the crystalline systems have been successfully realized in one-dimensional disordered atomic wires~\cite{Meier2018Science} and a photonic platform~\cite{Stutzer2018Nature}. On the other hand, the atomic quasicrystals may be grown by the deposition of atoms on the surfaces of quasicrystal substrates~\cite{jenks1998Langmuir,McGrath2002JPCM,Sharma2007AdvPhy,McGrath2012PTRSA,Huang2018PRL}, and the photonic quasicrystals have already been realized in experiments~\cite{Verbin2013PRL,Levi2011Science,Vardeny2013NatPho,Viebahn2019PRL,Steurer2007JPD}. Therefore, we suggest that the experimental realization of the TAI phase in the quasicrystals is promising in the atomic or photonic systems.

\section*{Acknowledgments}
B.Z. was supported by the National Natural Science Foundation of China (Grant No. 11274102), the Program for New Century Excellent Talents in University of Ministry of Education of China (Grant No. NCET-11-0960), and the Specialized Research Fund for the Doctoral Program of Higher Education of China (Grant No.
20134208110001). R.C. and D.-H.X. were supported by the NSFC (Grant No. 11704106) and the Scientific Research Project of Education Department of Hubei Province (Grant No. Q20171005). D.-H.X. also acknowledges the financial support of the Chutian Scholars Program in Hubei Province.

\end{document}